\documentclass[aps,pre,preprint,superscriptaddress,showpacs]{revtex4-1}
\usepackage{amsmath}
\usepackage{amssymb}
\usepackage{graphicx}

\begin{document}

\preprint{}

\title{A generalization of the cumulant expansion. Application to a scale-invariant probabilistic model}

\author{A. Rodr\'{\i}guez}
\affiliation{GISC, Dpto. de Matem\'{a}tica Aplicada y Estad\'{\i}stica,
        Universidad Polit\'{e}cnica de Madrid, Pza. Cardenal Cisneros s/n, 28040 Madrid, Spain}
\author{C. Tsallis}
\affiliation{Centro Brasileiro de Pesquisas F\'{\i}sicas and National Institute of Science and Technology for Complex Systems, Rua Xavier Sigaud 150, 22290-180 Rio de Janeiro, Brazil}
\affiliation{Santa Fe Institute, 1399 Hyde Park Road, Santa Fe, New Mexico 87501, USA}

\date{\today}

\begin{abstract}
As well known, cumulant expansion is an alternative way to moment expansion to fully characterize probability distributions provided {\em all} the moments exist. If this is not the case, the so called escort mean values (or $q-$moments) have been proposed to characterize probability densities with divergent moments [C. Tsallis {\em et al}, J. Math. Phys {\bf 50}, 043303 (2009)]. We introduce here a new mathematical object, namely the $q-${\em cumulants}, which, in analogy to the cumulants, provide an alternative characterization to that of the $q-$moments for the probability densities. To illustrate the technical details of the procedure, we apply this new scheme to further study a recently proposed family of scale-invariant discrete probabilistic models [A. Rodr\'{\i}guez {\em et al}, J. Stat. Mech. (2008) P09006; R. Hanel {\em et al}, Eur. Phys. J. B {\bf 72}, 263 (2009)] having $q-$Gaussians as limiting probability distributions.
\end{abstract}


\maketitle

\section{Introduction} 

In classical thermodynamics for short-range-interacting systems we have two types of thermodynamical quantities, namely the {\it extensive} (e.g., total energy, total entropy, volume, etc) and the {\it intensive} (e.g., temperature, pressure, chemical potential, etc) ones. Within the formalism of Boltzmann-Gibbs statistical mechanics, these quantities typically emerge as successive moments of the $N$-particle Hamiltonian. If, for whatever reason, exact calculations are not tractable, a variety of procedures exist which involve truncations at some order. It is desirable that such truncations maintain the extensive or intensive nature of the quantities. Excepting the first moments in terms of the total Hamiltonian, which always are extensive, all the  higher-order moments violate extensivity. There are however specific combinations of these moments which preserve the extensivity. These are the so-called cumulants. For example, the total specific heat, which has to be extensive, appears as a second-order cumulant, i.e., a convenient combination of first- and second-order moments. Naturally, cumulant expansions are mathematically legitimate {\it only when all moments are finite}. A variety of physical systems exist for which this property is not verified. The purpose of the present paper is to develop a generalized form of cumulant expansion which overcomes this restriction. These generalized cumulants are in turn based on consistently generalized moments ({\it escort moments} or {\it $Q$-moments}), which we shall introduce later on. In Section II we introduce the $q$-cumulants; in Section III we review a special family of scale-invariant probability models; in Section IV we study the corresponding $Q$-moments; in Section V we address the associated (unnormalized) $q$-cumulants. We summarize our results in Section~VI.

Let us briefly remind the standard cumulant expansion. For a probability density function $f(x)$ the cumulants $\kappa_j=g^{(j)}(0)$, $j=1,2,\dots$, are defined through the derivatives of the cumulant-generating function  \begin{equation}g(t)\equiv\ln M(t)=t\kappa_1+\frac{t^2}{2!}\kappa_2+\frac{t^3}{3!}\kappa_3+\cdots,\label{funcion_generadora_cumulantes}\end{equation} where 
\begin{equation}M(t)\equiv\langle e^{tx}\rangle=1+t\mu_1+\frac{t^2}{2!}\mu_2+\frac{t^3}{3!}\mu_3+\cdots\label{funcion_generadora_momentos}\end{equation} 
is the moment-generating function and  $\mu_j=\langle x^j\rangle=M^{(j)}(0)$, $j=0,1,2,\dots$, are the moments. 
Taking logarithms and Taylor expanding in Eq. \eqref{funcion_generadora_momentos} yields

\begin{align}\ln M(t)&=\frac{t}{1!}\mu_1\nonumber\\
&+\frac{t^2}{2!}(\mu_2-\mu_1^2)\nonumber\\
&+\frac{t^3}{3!}(\mu_3-3\mu_1\mu_2+2\mu_1^3)\nonumber\\
&+\frac{t^4}{4!}(\mu_4-4\mu_1\mu_3+12\mu_1^2\mu_2-3\mu_2^2-6\mu_1^4)\label{taylor_expansion}\\
&+\frac{t^5}{5!}(\mu_5-5\mu_1\mu_4-10\mu_2\mu_3+20\mu_1^2\mu_3-60\mu_1^3\mu_2+30\mu_1\mu_2^2+24\mu_1^5)\nonumber\\
&+\frac{t^6}{6!}(\mu_6-10\mu_3^2-15\mu_2\mu_4+30\mu_2^3-6\mu_1\mu_5+120\mu_1\mu_2\mu_3+30\mu_1^2\mu_4\nonumber\\&\quad\quad\;-270\mu_1^2\mu_2^2-120\mu_1^3\mu_3+360\mu_2\mu_4-120\mu_6)\nonumber\\
&\;\,\vdots\nonumber\end{align}
Comparing Eqs. \eqref{taylor_expansion} and \eqref{funcion_generadora_cumulantes} one gets $\kappa_1=\mu_1$, $\kappa_2=\mu_2-\mu_1^2$, $\kappa_3=\mu_3-3\mu_1\mu_2+2\mu_1^3,\dots$, which follow the general relation between moments and cumulants given by 
\begin{equation}\kappa_j=j!\sum_{\sum_{i=1}^j in_i=j}\frac{(-1)^{\bar n-1}}{\bar n}\binom{\bar n}{n_1\;n_2\;\quad n_j}\prod_{i=1}^j\left(\frac{\mu_i}{i!}\right)^{n_i};\quad j=1,2, \dots\label{cumulantes}\end{equation}
where the sum runs over solutions of the equation $\sum_{i=1}^j in_i=j$, with $n_i$, $i=1,2,\dots,j$ being nonnegative integers, $\bar n=\sum_{i=1}^j n_i$ and we have made use of the multinomial coefficient $\binom{\bar n}{n_1\;n_2\;\quad n_j}=\frac{\bar n!}{n_1!n_2!\cdots n_j!}$. Relation (\ref{cumulantes}) is a convenient re-writting of the relations given in \cite{Hanggi,Luciano}.
An alternative connection between cumulants and moments can be seen in \cite{risken}. 

As well known, the set of moments fully characterize a probability density 
function provided they are {\it all finite}, the set of cumulants (linear combinations of the moments) being an alternative and, for some purposes (see later on), more convenient description. Once the set of moments are known, the probability distribution  may be obtained via Fourier antitransforming the characteristic function $\varphi(t)=M(it)={\cal F}[f](t)=\int_{-\infty}^\infty dx e^{itx}f(x)$.  

Notice that although $\kappa_2=\langle(x-\langle x\rangle)^2\rangle$ and $\kappa_3=\langle(x-\langle x\rangle)^3\rangle$, in general the cumulants do {\em not} coincide with the centered moments since $\kappa_4\neq\langle(x-\langle x\rangle)^4\rangle$ and so on. 

Let us review now an important property. We assume $x=\xi_1+\xi_2+\cdots+\xi_N$, 
where $\{\xi_l\}$ are {\em any} $N$ {\em equal} and {\em independent} random variables. We straightforwardly verify that 
\begin{align}
\langle x\rangle &=N\langle\xi_1\rangle\nonumber\\
\langle x^2\rangle-\langle x\rangle^2&=N[\langle\xi_1^2\rangle-\langle\xi_1\rangle^2]\\
\langle x^3\rangle-3\langle x^2\rangle\langle x\rangle+2\langle x\rangle^3&=N[\langle\xi_1^3\rangle-3\langle\xi_1^2\rangle\langle\xi_1\rangle+2\langle\xi_1\rangle^3]\nonumber\\&\vdots\nonumber
\end{align}
In general
\begin{equation}\kappa_j(N)=N\kappa_j(1)\quad (j=1,2,3,\dots;\;N=1,2,3,\dots),\label{extensive}\end{equation}
where the notation $\kappa_j(N)$ is self-explanatory. In other words, {\em all} cumulants are 
{\em extensive} in the thermodynamical sense. The main purpose of the present paper is to discuss what happens with this property in the presence of  strong correlations such as those that are typical within $q-$statistics \cite{libro}, a current generalization of Boltzmann-Gibbs statistical mechanics (recovered as the $q=1$ particular instance). This generalization, sometimes referred to as {\it nonextensive statistical mechanics}, has received a wide variety of physical applications \cite{UpadhyayaRieuGlazierSawada2001,DanielsBeckBodenschatz2004,ArevaloGarcimartinMaza2007a,DouglasBergaminiRenzoni2006,LiuGoree2008,DeVoe2009,Borland2002a,Queiros2005,BurlagaVinas2005,BurlagaNess2009,BakarTirnakli2009,CarusoPluchinoLatoraVinciguerraRapisarda2007,CarvalhoSilvaNascimentoMedeiros2008,PickupCywinskiPappasFaragoFouquet2009,CMS2010}.

\section{$q-$cumulants}
Within the frame of $q-$statistics, a generalization of the concept of cumulant is in order. A first natural attempt for such generalization consists in just replacing the logarithm and exponential functions in \eqref{taylor_expansion} by the $q-$logarithm and $q-$exponential functions (see Appendix for definitions and corresponding expansions). Their respective Taylor expansions yields

\begin{align}
\ln_q^{}\langle e_q^{tx}\rangle&=
t\kappa_{1,q}+\frac{t^2}{2!}\kappa_{2,q}+\frac{t^3}{3!}\kappa_{3,q}+\frac{t^4}{4!}\kappa_{4,q}+\frac{t^5}{5!}\kappa_{5,q}+\cdots\nonumber\\
&=\frac{t}{1!}\mu_1\nonumber\\
&+\frac{t^2}{2!}q(\mu_2-\mu_1^2)\nonumber\\
&+\frac{t^3}{3!}q((2q-1)\mu_3-3q\mu_1\mu_2+(q+1)\mu_1^3)\nonumber\\
&+\frac{t^4}{4!}q((2q-1)(3q-2)\mu_4-4q(2q-1)\mu_1\mu_3+6q(q+1)\mu_1^2\mu_2-3q^2\mu_2^2-(q+1)(q+2)\mu_1^4)\nonumber\\
&+\frac{t^5}{5!}q((2q-1)(3q-2)(4q-3)\mu_5-5q(2q-1)(3q-2)\mu_1\mu_4-10q^2(2q-1)\mu_2\mu_3 \nonumber\\&\quad\quad\quad+10(q+1)q(2q-1)\mu_1^2\mu_3-10q(q+1)(q+2)\mu_1^3\mu_2 +15q^2(q+1)\mu_1\mu_2^2\nonumber
\\&\quad\quad\quad+(q+1)(q+2)(q+3)\mu_1^5)\nonumber
\\&\;\,\vdots\label{q-cumulants}
\end{align}
where we have defined the $q-${\em cumulant of order j}, $\kappa_{j,q}$, which is related with the moments through
\begin{equation}\kappa_{j,q}=j!\sum_{\sum_{i=1}^j in_i=j}\frac{(-1)^{\bar n-1}}{\bar n!}l(q,\bar n)\binom{\bar n}{n_1\;n_2\;\quad n_j}\prod_{i=1}^j\left(\frac{e(q,i)\mu_i}{i!}\right)^{n_i};\quad j=1,2, \dots\label{q-cumulantes}\end{equation} 
where $e(q,n)$ and $l(q,n)$ are the coefficients of the Taylor expansion of the $q-$exponential and $q-$logarithm functions respectively (see Appendix).

Notice that the moments $\{\mu_j\}$ in \eqref{q-cumulants} are exactly the same introduced in Eqs. \eqref{funcion_generadora_momentos} and \eqref{taylor_expansion}. In other words, $\{\mu_j\}$ do {\em not} depend on $q$. Notice also that $\kappa_{1,q}=\mu_1,\;\forall q$ (the $q-$cumulant of first order coincide with the first moment for any value of $q$), and $\kappa_{j,1}=\kappa_j\;\forall j$ (the $q-$cumulants reduce to the standard cumulants for $q=1$).

Although interesting in principle, we will show that the $q-$generalization which is useful is {\em not} the above one, but the one which follows now. We will define the {\em unnormalized} $q-$cumulants, $\kappa^\prime_{j,q}$, through

\begin{align}
\ln_q^{} M_q(t)&=
t\kappa^\prime_{1,q}+\frac{t^2}{2!}\kappa^\prime_{2,q}+\frac{t^3}{3!}\kappa^\prime_{3,q}+\frac{t^4}{4!}\kappa^\prime_{4,q}+\frac{t^5}{5!}\kappa^\prime_{5,q}+\cdots\nonumber\\
&=\frac{t}{1!}\mu^\prime_{1,q}\nonumber\\
&+\frac{t^2}{2!}q(\mu^\prime_{2,2q-1}-{\mu^\prime}^2_{1,q})\nonumber\\
&+\frac{t^3}{3!}q((2q-1)\mu^\prime_{3,3q-2}-3q\mu^\prime_{2,2q-1}\mu^\prime_{1,q}+(q+1){\mu^\prime}^3_{1,q})\nonumber\\
&+\frac{t^4}{4!}q((2q-1)(3q-2)\mu^\prime_{4,4q-3}-4q(2q-1)\mu^\prime_{3,3q-2}\mu^\prime_{1,q}+6q(q+1)\mu^\prime_{2,2q-1}{\mu^\prime}^2_{1,q}\nonumber\\&\quad\quad\quad-3q^2{\mu^\prime}^2_{2,2q-1}-(q+1)(q+2){\mu^\prime}^4_{1,q})\label{unnormalized-q-cumulants}\\
&+\frac{t^5}{5!}q((2q-1)(3q-2)(4q-3)\mu^\prime_{5,5q-4}-5q(2q-1)(3q-2)\mu^\prime_{4,4q-3}\mu^\prime_{1,q}\nonumber\\&\quad\quad\quad-10q^2(2q-1)\mu^\prime_{3,3q-2}\mu^\prime_{2,2q-1}\nonumber+10(q+1)q(2q-1)\mu^\prime_{3,3q-2}{\mu^\prime}^2_{1,q}\nonumber\\&\quad\quad\quad-10q(q+1)(q+2)\mu^\prime_{2,2q-1}{\mu^\prime}^3_{1,q}+15q^2(q+1){\mu^\prime}^2_{2,2q-1}\mu^\prime_{1,q}
\nonumber\\&\quad\quad\quad+(q+1)(q+2)(q+3){\mu^\prime}^5_{1,q})
\nonumber\\&\;\,\vdots\nonumber\end{align}
where we shall call $M_q(t)\equiv\langle e_q^{txf(x)^{q-1}}\rangle$ the {\em Q-moments generating function}. Expression \eqref{unnormalized-q-cumulants} makes use of the so called {\em unnormalized escort moments} \cite{TsallisPlastinoAlvarez-Estrada}, $\mu^\prime_{j,Q}\equiv\int dx\,x^jf(x)^Q$, related in turn to the standard {\em escort moments} (or $Q-${\em moments}),     
$\mu_{j,Q}=\mu^\prime_{j,Q}/\int dx\, f(x)^Q$, i. e., $x^j$ averaged over the {\em escort distribution} $f_Q(x)\equiv f(x)^Q/\gamma_Q$, with $\gamma_Q\equiv\int dx\, f(x)^Q$. In the limit $Q\to 1$, the escort distribution as well as the $Q-$moments, and the unnormalized $Q-$moments, tends to the standard ones. 

Notice from Eq.~\eqref{unnormalized-q-cumulants} that, for the first unnormalized $q-$cumulant we must use the $Q-$moment with $Q=q$ (in fact the unnormalized $q-$cumulant coincides with the unnormalized escort moment for $j=1$: $\kappa^\prime_{1,q}=\mu^\prime_{1,q}$), for the second unnormalized $q-$cumulant, $Q-$moments with two different values, namely $Q=q$ and $Q=2q-1$, must be used, and similarly for higher-order unnormalized $q$-cumulants. The relation between the unnormalized $q-$cumulants and the unnormalized $Q-$moments is thus given by
\begin{equation}\kappa^\prime_{j,q}=j!\sum_{\sum_{i=1}^j in_i=j}\frac{(-1)^{\bar n-1}}{\bar n!}l(q,\bar n)\binom{\bar n}{n_1\;n_2\;\quad n_j}\prod_{i=1}^j\left(\frac{e(q,i)\mu^\prime_{i,i(q-1)+1}}{i!}\right)^{n_i}\label{q-cumulantes-Q-momentos}
\end{equation} 

For $q=1$, the unnormalized escort moments appearing in Eqs.~\eqref{unnormalized-q-cumulants} and \eqref{q-cumulantes-Q-momentos}: $\mu^\prime_{i,Q}$, with $Q=i(q-1)+1$, $i=1,2,3,\dots$, reduce to the normalized escort moments which in turn coincide with the ordinary moments: $\mu^\prime_{i,1}=\mu_{i,1}=\mu_i\;\forall i$. The same happens with the unnormalized $q-$cumulants: $\kappa^\prime_{i,1}=\kappa_{i,1}=\kappa_i\;\forall i$, so Eqs. \eqref{cumulantes}, \eqref{q-cumulantes} and \eqref{q-cumulantes-Q-momentos} coincide. 

For the general $q\neq 1$ case, as shown in Ref. \cite{TsallisPlastinoAlvarez-Estrada}, the family of unnormalized escort moments  $\mu^\prime_{i,i(q-1)+1}$, $i=1,2,3,\dots$, provided they are {\it all finite}, may serve, together with their acompanying denominators, $\gamma_{i(q-1)+1}=\int dx\, f(x)^{i(q-1)+1}$, in the expresion of the corresponding escort moments, as an alternative description for the quite frequent
cases of probability density functions whose moments are not defined (e.g. moments of order $j\geqslant 2$ for $q-${\em Gaussians} with $q\geqslant 5/3$ \cite{Prato}). In analogy with the $Q=1$ case, once the $Q-$moments generating function is known, the probability distribution may be obtained via $q-$Fourier antitransforming the {\em q-characteristic function} $\varphi_q(t)=M_q(it)\equiv{\cal F}_q[f](t)=\int_{-\infty}^\infty dx\,e_q^{itxf(x)^{q-1}}f(x)$, $(f(x)\geqslant 0,\;q\geqslant 1)$, where we have made use of the so called  {\em q-Fourier transform} \cite{Umarov,Umarov2}, which is a (nonlinear) $q-$generalization of the standard Fourier transform.

\section{Scale Invariant Triangles}
\label{Scale_Invariant_Triangles}
We will apply the above formalism to further study a family of scale-invariant probabilistic models characterized by a real number $\nu>0$, first introduced in \cite{RodriguezSchwammleTsallis} and later generalized in \cite{HanelThurnerTsallis}. The model consists in a set of $N$ {\em equal}, {\em long-range-correlated binary} random variables $\{x_i\}$, which take values $0$ and $1$.

For the $2^N$ elementary events of the sample space of the $N$ variables system there is a set of only $(N+1)$ different probabilities given by \cite{HanelThurnerTsallis}
\begin{equation}r_{N,n}^{(\nu)}=\frac{B(n+\nu,N-n+\nu)}{B(\nu,\nu)}\label{triangulos_generalizados}\end{equation}
for $n=0,1,\dots,N$, where $B(x,y)$ stands for the Beta function. The $\{r_{N,n}^{(\nu)}\}$ satisfy the so called Leibnitz triangle rule \cite{RodriguezSchwammleTsallis}
\begin{equation}r_{N,n}^{(\nu)}+r_{N,n+1}^{(\nu)}=r_{N-1,n}^{(\nu)}\label{Leibnitz-rule}\end{equation}
which is a fingerprint of the scale invariant character of the system, which states that the marginal probability distribution of the $N$ system coincides with the joint probability distribution of the corresponding $(N-1)-$subsystem. In other words, in its continuous version, we have $\int p_N(x_1,x_2,\dots,x_N)dx_N=p_{N-1}(x_1,x_2,\dots,x_{N-1})$ (although similar in form, this conditions are not the Kolmogorov consistency conditions of a stochastic process ---see, for instance \cite{vanKampen}---. Indeed, the Kolmogorov conditions refer to a system of $N$ elements whereas here we impose nontrivial conditions connecting the joint probabilities of a $(N-1)-$subsystem with the marginal ones of a $N-$system).

For a fixed value of $\nu$, probabilities \eqref{triangulos_generalizados} may be displayed as a function of $N$ in a symmetric (i.e., $r_{N,n}^{(\nu)}=r_{N,N-n}^{(\nu)}$) triangle (thus the name {\it scale invariant triangles}). As an example, the $\nu=\frac{1}{2}$ triangle reads
\begin{center}
\begin{tabular}{ccccccccccccccccc}
&&&&&&&&$r_{N,n}^{(\frac{1}{2})}$&&&&&&&\\
&&&&&&&&&&&&&&&\\
$(N=0)\quad\quad$&&&&&&&&1&&&&&&&\\
$(N=1)\quad\quad$&&&&&&&$\dfrac{1}{2}$&&$\dfrac{1}{2}$&&&&&&\\
$(N=2)\quad\quad$&&&&&&$\dfrac{3}{8}$&&$\dfrac{1}{8}$&&$\dfrac{3}{8}$&&&&&\\
$(N=3)\quad\quad$&&&&&$\dfrac{5}{16}$&&$\dfrac{1}{16}$&&$\dfrac{1}{16}$&&$\dfrac{5}{16}$&&&&\\
$(N=4)\quad\quad$&&&&$\dfrac{35}{128}$&&$\dfrac{5}{128}$&&$\dfrac{3}{128}$&&$\dfrac{5}{128}$&&$\dfrac{35}{128}$&&&\\
$(N=5)\quad\quad$&&&$\dfrac{63}{256}$&&$\dfrac{7}{256}$&&$\dfrac{3}{256}$&&$\dfrac{7}{256}$&&$\dfrac{1}{256}$&&$\dfrac{63}{256}$&&\\
&&&&$\vdots$&&&&$\vdots$&&&&$\vdots$&
\end{tabular}
\end{center}

Due to the properties of the Beta function the coefficients of the triangle $r_{N,n}^{(\nu)}$ are rational numbers for any rational value of $\nu$. 

Special values of $\nu$ are $\nu=1$, for which the original Leibnitz triangle \cite{Leibnitz_book} is obtained $[1/r_{N,n}^{(1)}=(N+1)\binom{N}{n}]$, and the limiting case $r_{N,n}^{(\infty)} \equiv\lim_{\nu\to\infty} r^{(\nu)}_{N,n}=1/2^N$, for which the system becomes uncorrelated \cite{RodriguezSchwammleTsallis}. In the following, we shall refer to this limiting cases as the Leibnitz and the de Moivre-Laplace (or Boltzmann-Gibbs) limits respectively.

The probability distribution for the discrete variable $z=x_1+x_2+\dots+x_N$, which takes values $0,1,\dots,N$, is given by
\begin{equation}p_{N,n}^{(\nu)}\equiv P(z=N-n)=\binom{N}{n}r_{N,n}^{(\nu)}\label{probabilidades}\end{equation}
with $\sum_{n=0}^Np_{N,n}^{(\nu)}=1$, due to the degeneracy imposed by the identical character of de $N$ binary variables.
By properly centering and scaling with a map $n\to y(n)$, the distribution \eqref{probabilidades} is transformed to a new one, ${\cal P}^{(\nu)}(y)$, which tends to a $q-$Gaussian 
\begin{equation}{\cal P}^{(\nu)}(y)\to A_q^{}e_q^{-y^2}\label{limite}\end{equation}
for $N\to\infty$, with $A_q$ such $\int A_q^{}e_q^{-y^2}dy=1$, and a value of $q=q_\text{lim}$ which depends on $\nu$ and the chosen change of variable. The variable change, the intermediate distribution, and the value of $q_\text{lim}$ are given in Table \ref{tabla_1}. The Leibnitz case ($\nu=1$ hence $q \to -\infty$) corresponds to a uniform distribution while for the Boltzmann-Gibs case ($\nu=\infty$ hence $q=1$) the standard Gaussian is obtained.  
\begin{table}[h]
\begin{center}\begin{tabular}{c|c|c}
map&${\cal P}^{(\nu)}(y)$&$q_\text{lim}$\\\hline&&\\
$y=2\sqrt{\nu-1}\left(\dfrac{n}{N}-\dfrac{1}{2}\right)$&
$\dfrac{N}{2\sqrt{\nu-1}}p_{N,n}^{(\nu)}\;$&$\;q_\nu=\dfrac{\nu-2}{\nu-1}<1;\quad (\nu>1)$\\&&\\\hline &&\\
$y=\sqrt{\nu+\dfrac{1}{2}}\,\dfrac{\left(n-\dfrac{N}{2}\right)}{\sqrt{n(N-n)}}\;$&$\;\dfrac{4(n(N-n))^{3/2}}{N^2\sqrt{\nu+\dfrac{1}{2}}}p_{N,n}^{(\nu)}\;$&$\;\bar q_\nu=\dfrac{2\nu+3}{2\nu+1}>1;\quad (\nu>0)$
\end{tabular}\end{center}
\caption{The two different changes of variables, corresponding respectively to all real values $q_\text{lim}<1$   \cite{RodriguezSchwammleTsallis,HanelThurnerTsallis} and all real values $1<q_\text{lim}<3$ \cite{HanelThurnerTsallis}, and the associated the corresponding ${\cal P}^{(\nu)}(x)$ and $q_\text{lim}$. \label{tabla_1}}
\end{table}

\begin{figure}
\includegraphics[height=12cm=12cm,angle=-90,clip=]{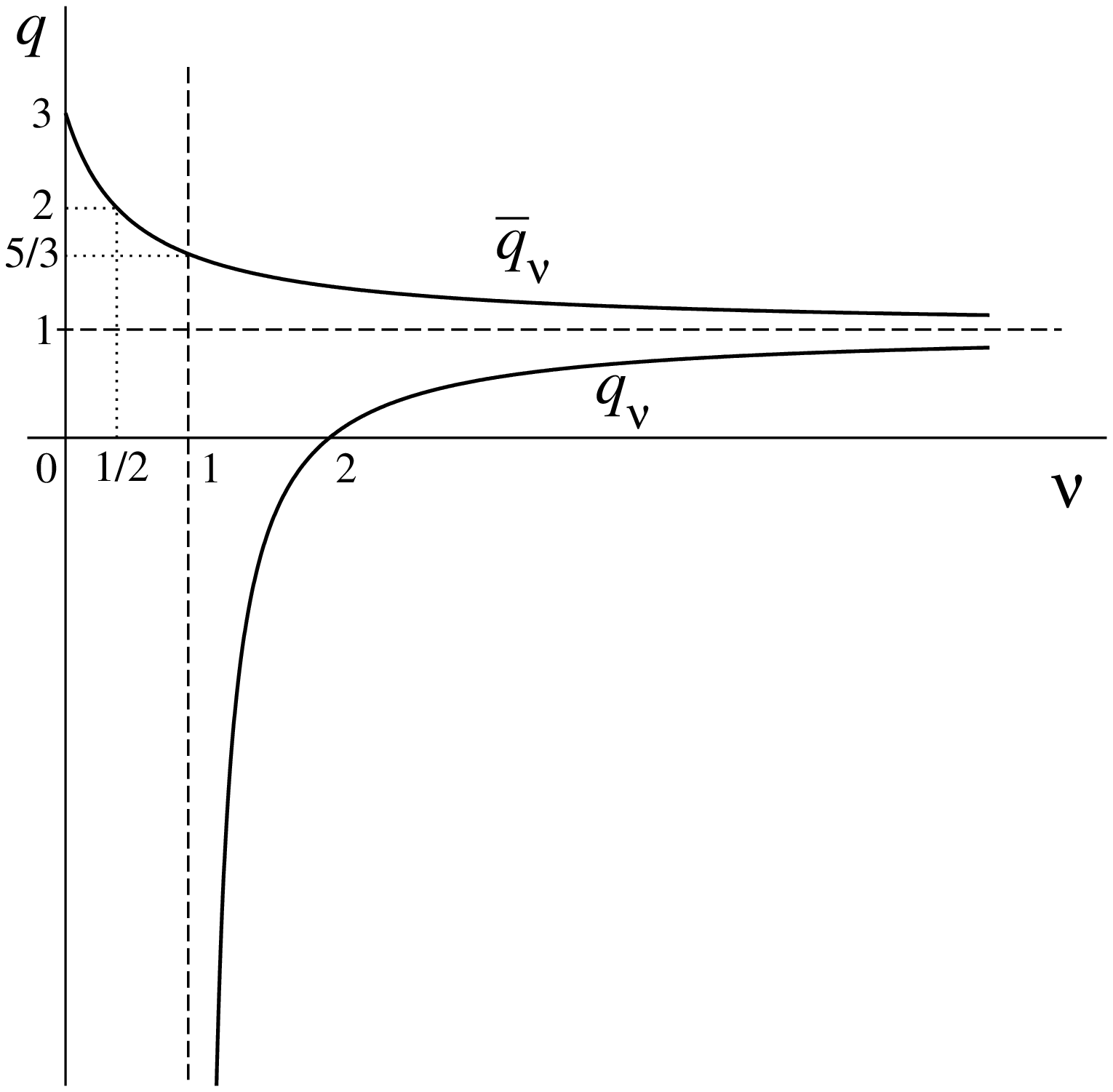}
\caption{Limiting values $q_\nu$ and $\bar q_\nu$ as a function of $\nu$. Notice the $\nu(q)$ is a single-valued function, whereas $q(\nu)$ is not.\label{fig_1}}
\end{figure}

Thus, depending on the chosen discretization (namely the Table I top or bottom discretization), for each value of $\nu\geqslant 1$ there exist two conjugated $q-$Gaussians, one of them with $1>q=q_\nu\in(-\infty,1)$ (with compact support $|x|<\frac{1}{\sqrt{1-q}}$) and the other with $1<q=\bar q_\nu\in(1,5/3]$ (the support being the whole real axis). In the case of $\nu$ being a positive integer, the different values of $q$ are shown in  Table \ref{tabla_2}. For $\nu\in(0,1)$, however, only one $q-$Gaussian with $1<q=\bar q_\nu\in (5/3,3)$ is obtained \cite{HanelThurnerTsallis}. Figure~\ref{fig_1} plots the two different $q_\text{lim}$ values as a function of $\nu$.
\begin{table}
\begin{center}\begin{tabular}{c|cccccccc}$\nu$&$\,0^+\,$&1/2&1&2&3&4&$\cdots$&$\infty$\\\hline$q_\nu$&&&$-\infty$&0&1/2&2/3&$\cdots$&$1^-$\\\hline$\bar q_\nu$&$\,3^-\,$&2&5/3&7/5&9/7&11/9&$\cdots$&$1^+$\\\hline\end{tabular}\end{center}
\caption{Values of $q$ as a function of $\nu$ for the two families of conjugated $q-$Gaussians.\label{tabla_2}}\end{table}

Relations $q(\nu)$ given in the third column of Table \ref{tabla_1} can be inverted so as to express the family of triangles as a function of $q$ \cite{HanelThurnerTsallis}:
\begin{equation}r_{N,n}^{(q)}=\left\{\begin{array}{lcc}\dfrac{B\left(n+\frac{q-2}{q-1},N-n+\frac{q-2}{q-1}\right)}{B\left(\frac{q-2}{q-1},\frac{q-2}{q-1}\right)};&&q<1\\\dfrac{1}{2^N};&&q=1\\&&\\\dfrac{B\left(n+\frac{3-q}{2(q-1)},N-n+\frac{3-q}{2(q-1)}\right)}{B\left(\frac{3-q}{2(q-1)},\frac{3-q}{2(q-1)}\right)};&&1<q<3\end{array}\right.\label{r(q)}\end{equation}
which allows as to fix a priori the value of $q$ and then get the triangle generating the corresponding $q-$Gaussian.

\begin{figure}
\includegraphics[width=8.6cm,clip=]{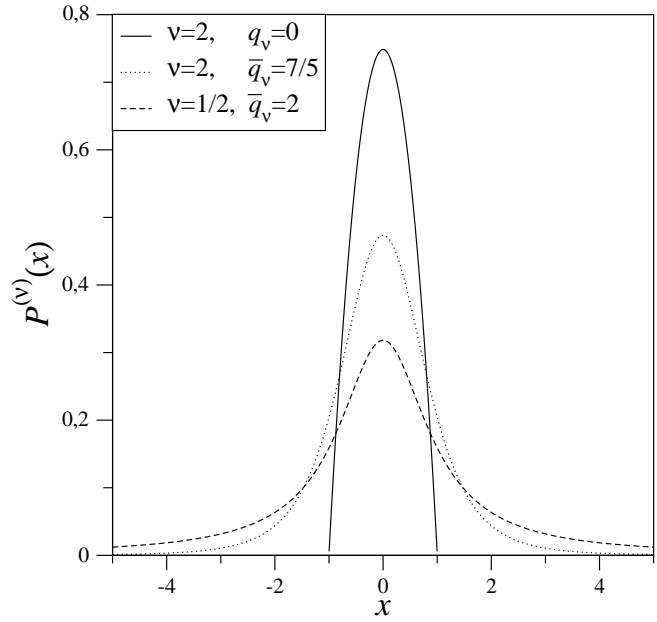}
\caption{Probability distribution ${\cal P}^{(\nu)}(x)$ for $\nu=2$, $q_\nu=0$ (solid line), $\nu=2$, $\bar q_\nu=7/5$ (dotted line), and $\nu=1/2$, $\bar q_\nu=2$ (dashed line). $N=1000$ in all cases\label{fig_2}.}
\end{figure}

\begin{figure}
\includegraphics[width=\linewidth,clip=]{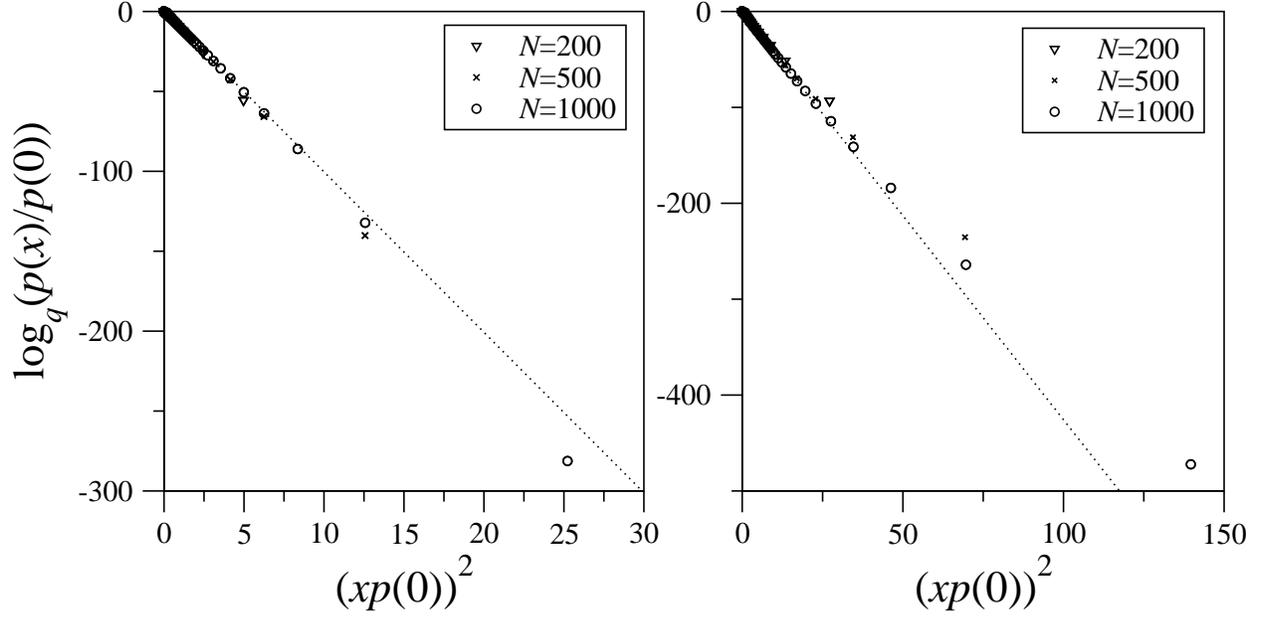}
\caption{$\log_{q_\nu}({\cal P}^{(\nu)}(x)/{\cal P}^{(\nu)}(0))$ versus $(x{\cal P}^{(\nu)}(0))^2$ for $\nu=1/2$, $\bar q_\nu=7/5$ (left) and $\nu=2$, $\bar q_\nu=2$ (right) for $N=200$ (triangles), 500 (crosses) and 1000 (circles). The dotted straight line is obtained by linear fitting after dropping the last four points.\label{fig_3}}
\end{figure}

Figure~\ref{fig_2} shows the probability distribution ${\cal P}^{(\nu)}(x)$ for a $N=1000$ system and two different values of the parameter $\nu$, namely $\nu=2$, for which $q_\nu=0<1$ and $\bar q_{\nu}=7/5>1$, and $\nu=1/2$, for which there exist one only value $\bar q_\nu=2$. The corresponding $q-$Gaussians (not shown) graphically overlap with the shown distributions. To further check Eq. \eqref{limite}, Fig.~\ref{fig_3} shows $\log_{q_\nu}({\cal P}^{(\nu)}(x)/{\cal P}^{(\nu)}(0))$ versus $(x{\cal P}^{(\nu)}(0))^2$ --- which, in the $N\to\infty$ limit, coincides with a straight line for a $q-$Gaussian --- for $\nu=1/2$, $\bar q_\nu=7/5$ (left) and $\nu=2$, $\bar q_\nu=2$ (right) for $N=200$, 500 and 1000. As can be seen in the figure, points far from the origin (corresponding to small values $n=0,1,\dots$, and their symmetric $n=N-1, N-2,\dots$) deviate from the fitting to a straight line. Nevertheless this is a finite size effect since increasing $N$ has the effect of bringing them back to the line. For values of $\nu$ and discretizations corresponding to compact support $q-$Gaussians (not shown) this finite size effect is not present and the fitting is improved.   

\section{Scaling of the $Q-$moments}

Our aim in this section is to study the scaling with the system size of the $Q-$moments and  $q-$cumulants of the family of triangles \eqref{triangulos_generalizados} or, equivalently \eqref{r(q)}. 

For our model, the escort moments of interest read
\begin{equation}\mu_{j,Q}^{(\nu)}\equiv\langle y^j\rangle_Q=\dfrac{\displaystyle\sum_{n=0}^Nn^j\left(p_{N,n}^{(\nu)}\right)^Q}{\displaystyle\sum_{n=0}^N\left(p_{N,n}^{(\nu)}\right)^Q}=\dfrac{{\mu^\prime}^{(\nu)}_{j,Q}}{\gamma_Q^{(\nu)}}\label{momentos_escort}\end{equation} 
where the numerator of the last fraction stands for the corresponding unnormalized $Q-$moment, the denominator being the normalization factor of the escort distribution. 

Straightforward calculations based on the symmetric character of triangles \eqref{triangulos_generalizados} allows us to exactly calculate the first-order escort moments as \begin{equation}\mu_{1,Q}^{(\nu)}=\kappa_{1,Q}^{(\nu)}=\frac{N}{2} \,,\quad\forall Q,\;\forall \nu.\label{j=1}\end{equation} 
The higher-order escort moments for the aforementioned Boltzmann-Gibbs and Leibnitz limiting cases are addressed next. 

\subsection{Boltzmann-Gibbs limit}
To begin with, we shall study the standard $(Q=1)$ moments for the Boltzmann-Gibbs ($\nu\to\infty$) limit. In this case the probability distribution \eqref{probabilidades} reduces to the binomial distribution $p_{N,n}^{(\infty)}=\frac{1}{2^N}\binom{N}{n}$ (which tends to the Gaussian distribution in the thermodynamic limit $N\to\infty$). Taylor expanding its moment generating function $M(t)=\langle e^{tx}\rangle=\frac{1}{2^N}\sum_{n=0}^N\binom{N}{n}e^{tn}=((1+e^t)/2)^N$, the standard moments $\mu_{j,1}^{(\infty)}=M^{(j)}(0)$ of the binomial distribution are easily obtained: 
\begin{align}\mu_{1,1}^{(\infty)}&=\frac{N}{2}\nonumber\\
           \mu_{2,1}^{(\infty)}&=\frac{N(N+1)}{4}\nonumber\\
	   \mu_{3,1}^{(\infty)}&=\frac{N^2(N+3)}{8} \label{momentos_binomial}\\
	   \mu_{4,1}^{(\infty)}&=\frac{N(N^3+6N^2+3N-2)}{16}\nonumber\\
	   \mu_{5,1}^{(\infty)}&=\frac{N(N^4+10N^3+15N^2-10N)}{32}\nonumber\\
	     &\;\;\vdots\nonumber\end{align}

Though not shown here, it is worthy noticing the unexpected result that the scaling law \eqref{momentos_binomial} holds for the escort moments of the binomial distribution for {\it any value of $Q$}, that is for the binomial distribution one gets
\begin{equation}\mu_{j,Q}^{(\infty)}\sim \frac{1}{2^j}N^j\,, \quad\forall Q\end{equation}

On the other hand, the standard ($q=1$) cumulants $\kappa_{j,1}^{(\infty)}$ of the binomial distribution can be obtained making use of Eq.~ \eqref{cumulantes} or, alternatively, as the derivatives of the cumulant generating function $g(t)=N\ln((1+e^t)/2)$: 
\begin{equation}\kappa_{1,1}^{(\infty)}=\frac{N}{2},\quad\kappa_{2,1}^{(\infty)}=\frac{N}{4},\quad\kappa_{3,1}^{(\infty)}=0,\quad\kappa_{4,1}^{(\infty)}=-\frac{N}{8},\quad\kappa_{5,1}^{(\infty)}=0,\quad\dots\label{cumulantes_Boltzmann}\end{equation}
where all the odd cumulants vanish from the third on. It has to be noticed that, for $N\to\infty$, while the moments follow the rule $\mu_{j,1}^{(\infty)}\propto N^j$, the cumulants remain extensive, that is, $\kappa_{j,1}^{(\infty)}\propto N$. In fact, we easily verify that the results in Eq. \eqref{cumulantes_Boltzmann} satisfy Eq. \eqref{extensive}.

\subsection{Leibnitz limit}
In the specially simple Leibnitz case ($\nu=1$), the probability distribution \eqref{probabilidades} reduces to the uniform one $p_{N,n}^{(1)}=\frac{1}{N+1}$, which once introduced in Eq.~\eqref{momentos_escort} yields $\mu_{j,Q}^{(1)}=\frac{1}{N+1}\sum_{n=0}^Nn^j$ {\it for all $Q$}, so the escort moments are independent of the value of $Q$, thus coinciding with the standard ($Q=1$) moments, which may again be calculated as the derivatives of the corresponding moment generating function $M(t)=\langle e^{tx}\rangle=\frac{1}{N+1}\sum_{n=0}^Ne^{tn}=\frac{1}{N+1}\frac{1-e^{(N+1)t}}{1-e^t}$, $M(0)=1$:
\begin{align}
\mu_{1,Q}^{(1)}&=\dfrac{N}{2}\nonumber\\
\mu_{2,Q}^{(1)}&=\dfrac{N^2}{3}+\dfrac{N}{6}\nonumber\\
\mu_{3,Q}^{(1)}&=\dfrac{N^3}{4}+\dfrac{N^2}{4}\label{momentos_Leibnitz}\\
\mu_{4,Q}^{(1)}&=\dfrac{N^4}{5}+3\dfrac{N^3}{10}+\dfrac{N^2}{30}-\dfrac{N}{30}\nonumber\\
\mu_{5,Q}^{(1)}&=\dfrac{N^5}{6}+\dfrac{N^4}{3}+\dfrac{N^3}{12}-\dfrac{N^2}{12}\nonumber\\
 &\;\;\vdots\nonumber\end{align}

In turn, the standard ($q=1$) cumulants $\kappa_{j,1}^{(1)}$ of the discrete uniform distribution can be obtained through Eq. \eqref{cumulantes} or from the cumulant generating function $g(t)=\ln\left(\frac{1}{N+1}\frac{1-e^{(N+1)t}}{1-e^t}\right)$, $g(0)=0$:
\begin{align}\kappa_{1,1}^{(1)}&=\frac{N}{2}\nonumber\\
              \kappa_{2,1}^{(1)}&=\frac{N^2}{12}+\frac{N}{6}\nonumber\\
	      \kappa_{3,1}^{(1)}&=0\label{cumulantes_Leibnitz}\\	      \kappa_{4,1}^{(1)}&=-\frac{N^4}{120}-\frac{N^3}{30}-\frac{N^2}{20}
-\frac{N}{30}\nonumber\\ \kappa_{5,1}^{(1)}&=0\nonumber\\&\;\;\vdots\nonumber\end{align}
where again the odd cumulants of order greater than one vanish. Scaling of cumulants \eqref{cumulantes_Leibnitz} may be expressed as $\kappa_{j,1}^{(1)}\sim\frac{B_j}{j}N^j$, $B_j$ being the Bernoulli numbers ($B_1=1/2$, $B_2=1/6$, $B_3=0$, $B_4=-1/30$, $B_5=0$, $B_6=1/42,\dots$).

\subsection{General case}

We shall now allow $\nu$ to take on any positive value in \eqref{momentos_escort}.
In principle, $Q$ may take on any real value. However, as we will be interested in the escort moments involved in the expression of the unnormalized $q-$cumulants \eqref{q-cumulantes-Q-momentos}, we will first focus on the case $Q=j^\prime(q-1)+1$ (the extra index $j^\prime$ has been introduced in order to allow for a more general description though the only escort moments appearing in Eq. \eqref{q-cumulantes-Q-momentos} are those with $j^\prime=j$). Furthermore, for the escort moment of the $\nu$ triangle we will take either $q=q_\nu$ or $q=\bar q_\nu$ as given in Table \ref{tabla_1}.  

\begin{figure}
\includegraphics[height=\linewidth,angle=-90,clip=]{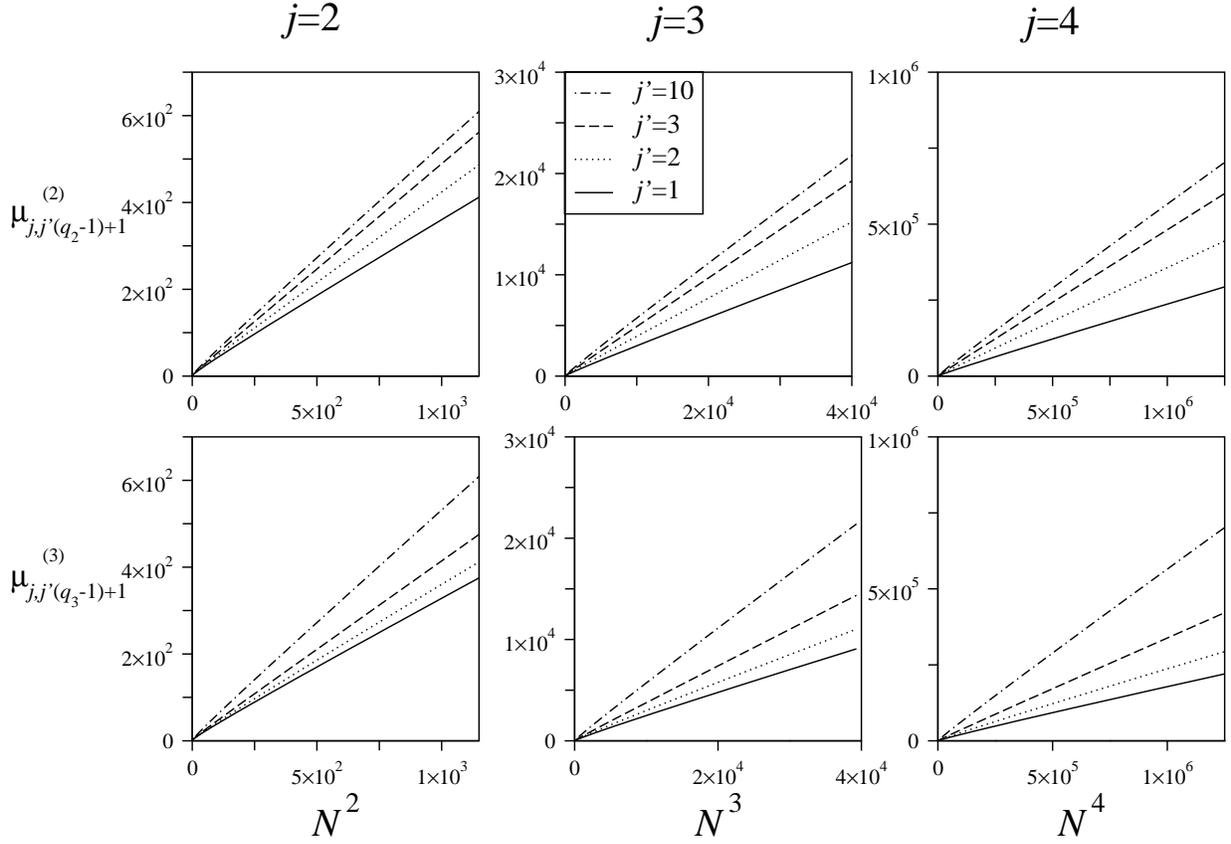}
\caption{Scaling of the escort moments \eqref{momentos_escort} with $Q=j^\prime(q_\nu-1)+1$ for different values of $j^\prime$, for  $\nu=2$ (up), $\nu=3$ (down) and $j=2$ (left), $j=3$ (center) and $j=4$ (right). The behavior for $q=\bar q_\nu$ is totally analogous.\label{fig_4} For the not shown $j=1$ case all the curves collapse in one: $\mu^{(\nu)}_{1,j^\prime(q_\nu-1)+1}=\frac{N}{2}$ since relation \eqref{j=1} is satisfied.}  
\end{figure}

Figure~\ref{fig_4} shows the scaling of the escort moments $\mu_{j,j^\prime(q_\nu-1)+1}^{(\nu)}$ of orders $j=2, 3$ and 4, for $\nu=2$ and 3 and different values of $j^\prime$. The following trend is observed 
\begin{equation}\mu_{j,j^\prime(q_\nu-1)+1}^{(\nu)}\sim A_{j,j^\prime, q_\nu}N^j;\quad\forall\nu,\,\forall j^\prime\label{general}\end{equation}
so the exponent of the scaling of the escort moments depends {\em only} on the order $j$ of the moment not on $j^\prime$, neither on $q_\nu$, which considerably simplifies the discussion. From now on we will restrict ourselves to the case $j^\prime=j$ (and correspondingly denote $A_{j,q_\nu}\equiv A_{j,j^\prime=j, q_\nu}$ in \eqref{general}). Not shown results reveal that Eq.~\eqref{general} is also valid for $Q=j^\prime(q-1)+1$ with $q=\bar q_\nu$.

\begin{figure}
\includegraphics[height=\linewidth,angle=-90,clip=]{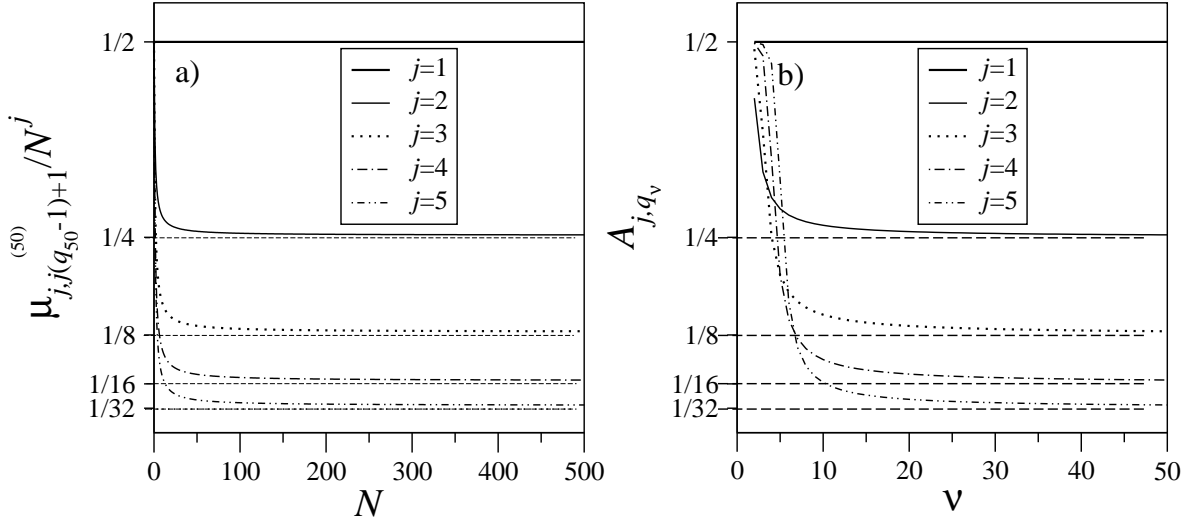}
\caption{a) Ratio $\mu_{j,j(q_\nu-1)+1}^{(\nu)}/N^j$ with $\nu=50$ as a function of $N$ and $j=1,2,3,4$ and 5 from top to bottom. b) $A_{j,q_\nu}$, calculated as the ratio $\mu_{j,j(q_\nu-1)+1}^{(\nu)}/N^j$ with $N=500$, as a function of $\nu$ and the same values of $j$ as in a).   
\label{fig_5}}
\end{figure}

The prefactor $A_{j,q_\nu}$ in the saling law \eqref{general} has already been analitically obtained in the Leibnitz ($q=-\infty$) case  as $A_{j,-\infty}=\frac{1}{j+1}$ as well as in the in the opposite limit, the Boltzmann-Gibbs $(q=1)$ case as $A_{j,1}=\frac{1}{2^j}$ (see Eqs. \eqref{momentos_Leibnitz} and \eqref{momentos_binomial} respectively). A transition between both behaviours is expected for the limit $q\to 1^\pm$. In order to numerically obtain $A_{j,q_\nu}$, 
Fig.~\ref{fig_5}a shows the ratio $\mu_{j,j(q_\nu-1)+1}^{(\nu)}/N^j$ for $\nu=50$ as a function of $N$ for different values of $j$. It is observed that $A_{j,q_{50}}\sim\frac{1}{2^j}$. Figure~\ref{fig_5}b shows again $A_{j,q_\nu}$, now approximated as the ratio $\mu_{j,j(q_\nu-1)+1}^{(\nu)}/N^j$ for $N=500$ as a function of $\nu$. It is clearly observed the expected asymptotic limit
\begin{equation}\lim_{\nu\to\infty}A_{j,q_\nu}=\frac{1}{2^j}\label{prefactor}\end{equation}

A further generalization for the scaling of the $Q-$moments must be done. Not shown calculations reveal that relation \eqref{general} still holds when $Q=j(q-1)+1$ and $q$ takes on {\em any} real value, not necessarily the values indicated in Table \ref{tabla_1}, (which in turn implies that $Q$ may take on any real value). Thus, it can be stated that  
\begin{equation}\mu_{j,j(q-1)+1}^{(\nu)}\sim A_{j,q,\nu}N^j;\quad\forall q,\;\forall \nu\label{mas_general}\end{equation}
which means that the whole family of triangles \eqref{triangulos_generalizados}, or equivalently \eqref{r(q)} share the same scaling behaviour for their escort moments $\mu_{j,Q}^{(\nu)}$.

\section{Scaling of the unnormalized $q-$cumulants}
\label{scaling_q-cumulants}
We shall now turn to the study of the unnormalized $q-$cumulants $\kappa^\prime_{j,q}$ given in Eq.~\eqref{q-cumulantes-Q-momentos}. Insomuch the unnormalized $q-$cumulant of order $j$, $\kappa^\prime_{j,q}$, is a linear combination of products of unnormalized $Q-$moments, which is homogeneous in the order $j$ of the $q-$cumulant (see restriction $\sum_{i=1}^j in_i=j$ in the sum defining the unnormalized  $q-$cumulant in Eq.~\eqref{q-cumulantes-Q-momentos}), the scaling of $\kappa^\prime_{j,q}$ will coincide with the scaling of the unnormalized $Q-$moment $\mu^\prime_{j,Q}$ with $Q=j(q-1)+1$.

\begin{figure}
\includegraphics[width=\linewidth,clip=]{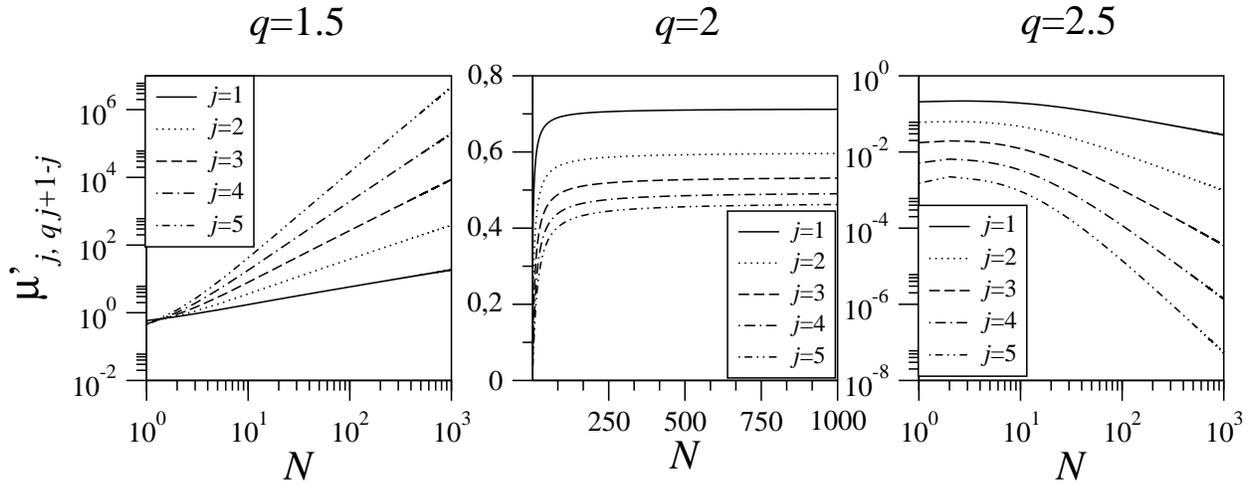}
\caption{Scaling of the unnormalized escort moments ${\mu^\prime}^{(\nu)}_{j,j(q-1)+1}$ for the triangle $\nu=3$ ($q_3=1/2$, $\bar q_3=9/7$) for $q=3/2$ (left) 2 (center) and 5/2 (right) for $j=1$ to $5$. The scaling law \eqref{escalado_Q_momentos_no_normalizados} is obtained. 
\label{fig_6}}
\end{figure}

Figure~\ref{fig_6} shows the asymptotic behaviour of the unnormalized escort moments ${\mu^\prime}^{(\nu)}_{j,j(q-1)+1}$ for the $\nu=3$ ($q_3=1/2$, $\bar q_3=9/7$) triangle for $q=3/2$ (left) 2 (center) and 5/2 (right). The following scaling relation is fulfilled 
\begin{equation}{\mu^\prime}^{(\nu)}_{j,j(q-1)+1}\sim A^\prime_{j,q,\nu}N^{j(2-q)}\label{escalado_Q_momentos_no_normalizados}\end{equation} 

Thus, the independence of the value of $q$ characterizing the scaling exponent of the escort moments (see Eq. \eqref{mas_general}) no longer holds for the unnormalized escort moments. Some comments must be made on the validity of scaling law \eqref{escalado_Q_momentos_no_normalizados}. It is valid for any positive $q$ for $\nu\geqslant 1$ (though it may fail for $q$ close to 0 and values of $j>1$). In the case $\nu<1$, Eq.~\eqref{escalado_Q_momentos_no_normalizados} holds only for $q\leqslant 1$, while the exponent scaling depends on the value of $\nu<1$ in a complicated fashion for $q>1$. It must be said, however, that relation \eqref{escalado_Q_momentos_no_normalizados} is only approximated when $q$ approaches 2. For $q=2$, in the transition from the increasing to the decreasing trend of ${\mu^\prime}^{(\nu)}_{j,Q}$ with $N$, a slight increase of the unnormalized $Q-$moments with $N$ is observed (see Fig.~\ref{fig_6}) instead of the expected constant value. We will discuss this point in detail later when studying the scaling of the unnormalized $q-$cumulants. Finally, for $q=1$ the unnormalized moments reduce to the standard ones and relations \eqref{mas_general} and \eqref{escalado_Q_momentos_no_normalizados} coincide.  

By comparing Eqs. \eqref{momentos_escort}, \eqref{mas_general} and \eqref{escalado_Q_momentos_no_normalizados}, it is readily deduced the scaling law for the normalizing coefficients of the escort distributions $\gamma_Q^{(\nu)}$ as
\begin{equation}\gamma^{(\nu)}_{j(q-1)+1}\sim 	\frac{A^\prime_{j,q,\nu}}{A_{j,q,\nu}}N^{j(1-q)}\label{escalado_coefficients}\end{equation}
within the same range of parameters as stated above. For the $q=1$ case, Eq.~\eqref{escalado_coefficients} is trivially fulfilled, representing the normalization of the family of probability distributions \eqref{probabilidades} as $\gamma_1^{(\nu)}=\sum_{n=0}^Np_{N,n}^{(\nu)}=1$.

As already mentioned, an analogous relation to \eqref{escalado_Q_momentos_no_normalizados} holds for the unnormalized $q-$cumulants
\begin{equation}{\kappa^\prime}^{(\nu)}_{j,q}\sim C_{j,q,\nu}N^{j(2-q)}\label{escalado_q-cumulantes_no_normalizados}\end{equation} 
with the same restrictions for the values of $q$ and $\nu$. Special cases are $q=0$, for which {\em all} the unnormalized $0-$cumulants vanish except for the first one (see Eq.~ \eqref{unnormalized-q-cumulants}), which coincides with the corresponding unnormalized escort moment ${\kappa^\prime}^{(\nu)}_{1,0}={\mu^\prime}^{(\nu)}_{1,0}\propto N^2$, and $q=1$, for which the standard cumulants are recovered as well as an analogous scaling law as that for the Leibnitz case \eqref{cumulantes_Leibnitz} for any value of $\nu$, that is, $\kappa_{j,1}^{(\nu)}\propto N^j$ with vanishing odd cumulants from the third on.

\begin{figure}
\centering\includegraphics[width=.5\linewidth,clip=]{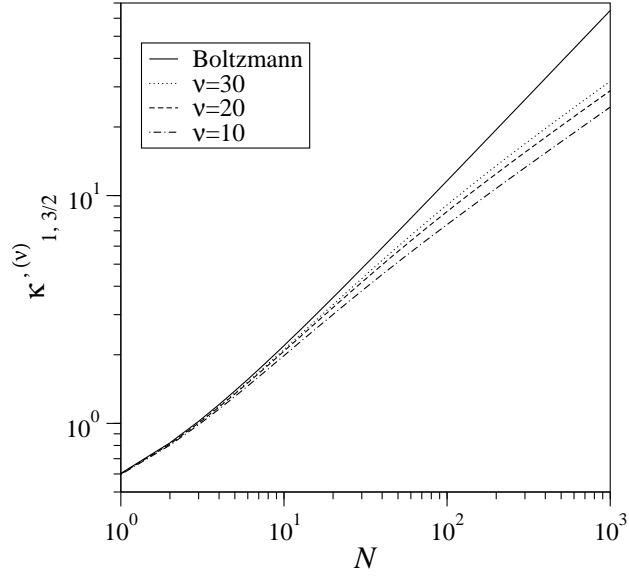}
\caption{Scaling of ${\kappa^\prime}^{(\nu)}_{1,3/2}={\mu^\prime}^{(\nu)}_{1,3/2}$  for $\nu=10$, 20, 30 and $\infty$. The  scaling exponent is 1/2 for $\nu<\infty$ and 3/4 for $\nu=\infty$.
\label{fig_7}}
\end{figure}

Nevertheless, relations \eqref{escalado_Q_momentos_no_normalizados} to \eqref{escalado_q-cumulantes_no_normalizados} are no longer valid in the $\nu\to\infty$ limit, that is, for the binomial distribution. It may be shown that in this case the scaling for $N\to\infty$ is
\begin{equation}{\mu^\prime}^{(\infty)}_{j,j(q-1)+1}\propto{\kappa^\prime}^{(\infty)}_{j,q}\propto N^{j(3-q)/2};\quad \gamma^{(\infty)}_{j(q-1)+1}\propto N^{j(1-q)/2} \,,\label{scaling_Boltzmann}\end{equation}
which coincide with relations \eqref{escalado_Q_momentos_no_normalizados} to \eqref{escalado_q-cumulantes_no_normalizados} for $q=1$. The very slow and nonuniform character of the convergence in $\lim_{\nu\to\infty}r^{(\nu)}_{N,n}=1/2^N$ and the numerical difficulties in dealing with coefficients $r^{(\nu)}_{N,n}$ for large $\nu$ and $N$ make it subtle to study in more detail the transition of the scaling exponents from the $\nu<\infty$ to the Boltzmann-Gibbs case. Figure~\ref{fig_7} shows the scaling of the first unnormalized $3/2-$cumulant for triangles with $\nu=10$, 20 and 30, together with the Boltzmann-Gibbs case. The crossover from the scaling exponent 1/2 given by \eqref{escalado_q-cumulantes_no_normalizados} for $\nu<\infty$ and exponent $3/4$ indicated in  \eqref{scaling_Boltzmann} for $\nu=\infty$ is clearly observed.         
Finally, Eq.~\eqref{scaling_Boltzmann} is valid for $q\geqslant 1$ for the unnormalized escort moments ${\mu^\prime}^{(\infty)}_{j,j(q-1)+1}$ and their corresponding normalizing coefficients $\gamma^{(\infty)}_{j(q-1)+1}$. However, for the unnormalized $q-$cumulants ${\kappa^\prime}^{(\infty)}_{j,q}$ its validity restricts to $q>1$. The case $q=1$ must be excluded since the  $1-$cumulants (normalized or not) coincide with the standard ones whose scaling $\kappa^{(\infty)}_{j,1}\propto N$ has been already obtained in \eqref{cumulantes_Boltzmann}. 

Let us connect now the values of $\nu$ and $q$ in Eq.~\eqref{escalado_q-cumulantes_no_normalizados}, i. e., let us study, for the $\nu$ triangle \eqref{triangulos_generalizados}, the $q-$cumulants with $q$ being the corresponding limiting value of the $q-$Gaussian \eqref{limite}. As stated in Table \ref{tabla_1}, there is only one limiting value, $q_\text{lim}=\bar q_\nu$ for $\nu\in(0,1)$, but two different options, namely $q_\text{lim}=q_\nu$ and $q_\text{lim}=\bar q_\nu$, for $\nu\geqslant 1$. We shall denote with $q_{(\nu)}$ any of them and calculate the unnormalized $q-$cumulant ${\kappa^\prime}^{(\nu)}_{j,q_{(\nu)}}$. 

Following the comments made for the validity of Eqs. \eqref{escalado_Q_momentos_no_normalizados} to \eqref{escalado_q-cumulantes_no_normalizados}, we may substitute $q$ by $q_{(\nu)}$ in Eq.~\eqref{escalado_q-cumulantes_no_normalizados} if $\nu\geqslant 1$ since it is valid for any value of $q$. The $\nu<1$ case must be studied separately since for this case $q_{(\nu)}=\bar q_\nu>1$ and, as already mentioned, Eq.~\eqref{escalado_q-cumulantes_no_normalizados} no longer holds. Nevertheless, provided $1/2<\nu<1$, Eq.~\eqref{escalado_q-cumulantes_no_normalizados} still holds but now only for the specific value $q=q_{(\nu)}$. On the contrary, for $\nu<1/2$ (hence $\bar q_\nu>2$), the scaling exponent reverts its sign. Sumarizing, the following scaling law is obtained \begin{equation}{\kappa^\prime}^{(\nu)}_{j,q_{(\nu)}}\sim C_{j,q_{(\nu)}}N^{j|2-q_{(\nu)}|}\label{escalado_q-cumulantes_no_normalizados_q_nu}\end{equation} 

\begin{figure}
\centering\includegraphics[height=\linewidth,angle=-90,clip=]{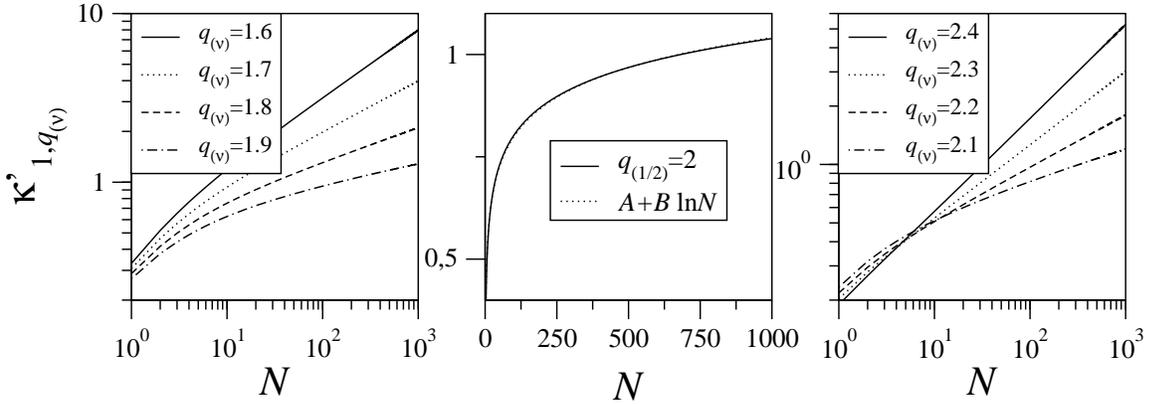}
\caption{Log-log plots of ${\kappa^\prime}^{(\nu)}_{1,q_{(\nu)}}={\mu^\prime}^{(\nu)}_{1,q_{(\nu)}}$ versus $N$ for $q_{(\nu)}=\bar q_\nu=1\text{.}6$ to $\bar q_\nu=1\text{.}9$ (left) and $\bar q_\nu=2\text{.}1$ to $\bar q_\nu=2\text{.}4$ (right). The center panel shows the $\bar q_{\frac{1}{2}}=2$ case and its fitting to a $A+B\ln N$ curve with $A=0\text{.}315$ and $B=0\text{.}105$. Both curves virtually overlap.
\label{fig_8}}
\end{figure}

\begin{figure}
\centering\includegraphics[width=8.6cm,clip=]{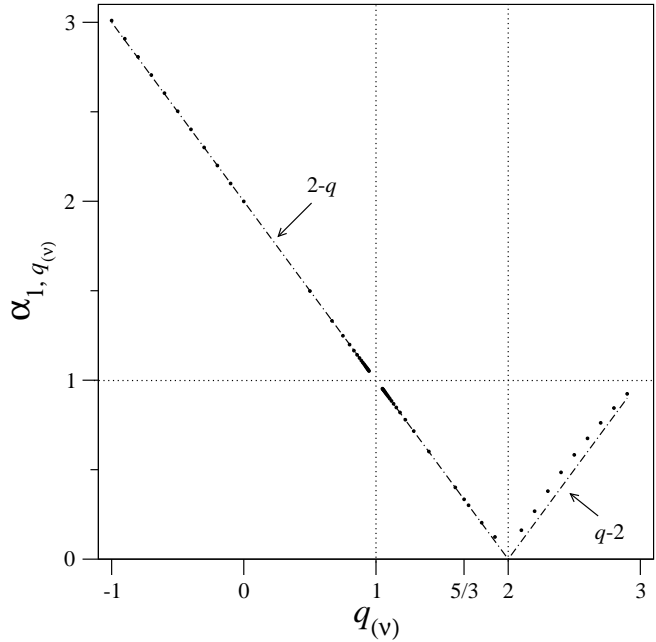}
\caption{Scaling exponent $\alpha_{1,q_{(\nu)}}$ of the first order unnormalized $q-$cumulant ${\kappa^\prime}^{(\nu)}_{1,q_{(\nu)}}={\mu^\prime}^{(\nu)}_{1,q_{(\nu)}}\sim C_{1,q_{(\nu)}}N^{\alpha_{1,q_{(\nu)}}}$. A small deviation from the scaling law $\alpha_{1,q_{(\nu)}}=|2-q_{(\nu)}|$ given in Eq.~\eqref{escalado_q-cumulantes_no_normalizados_q_nu} is observed for $q_{(\nu)}>2$. Exponents have been obtained via linear regresion of the $\ln({\kappa^\prime}^{(\nu)}_{1,q_{(\nu)}})$ versus $\ln N$ curves for $N$ from 500 to 1000. The $|q-2|$ curve is also plotted as a guide to the eye.
\label{fig_9}}
\end{figure}

Some comments on the validity of relation \eqref{escalado_q-cumulantes_no_normalizados_q_nu} must be made. As shown in Fig.~\ref{fig_8}, for $\bar q_\frac{1}{2}=2$, a logarithmic increase with $N$ of ${\kappa^\prime}^{(\frac{1}{2})}_{1,2}={\mu^\prime}^{(\frac{1}{2})}_{1,2}$ is obtained instead of the expected constant value predicted by  \eqref{escalado_q-cumulantes_no_normalizados_q_nu}, which therefore is no longer valid in this case and has to be replaced by a logarithmic law ${\kappa^\prime}^{(\frac{1}{2})}_{1,2}\sim A+B\ln N$ whose coefficients are found to be $A=0\text{.}315$ and $B=0\text{.}105$. This logarithmic correction influences numerically the scaling relation \eqref{escalado_q-cumulantes_no_normalizados_q_nu} for $q_{(\nu)}$ close to 2, as can be seen in Fig.~\ref{fig_8}. Finally, for the $q<0$ case, the scaling law still works \eqref{escalado_q-cumulantes_no_normalizados_q_nu} for $j=1$. Figure~\ref{fig_9} shows the actual value of the scaling exponent $\alpha_{1,q_{(\nu)}}=|2-q_{(\nu)}|$ of the first order unnormalized $q-$cumulant ${\kappa^\prime}^{(\nu)}_{1,q_{(\nu)}}={\mu^\prime}^{(\nu)}_{1,q_{(\nu)}}\sim C_{1,q_{(\nu)}} N^{\alpha_{1,q_{(\nu)}}}$. A small deviation from the scaling law \eqref{escalado_q-cumulantes_no_normalizados_q_nu} is observed for $q_{(\nu)}>2$.

Concerning the proportionality coefficient in \eqref{escalado_q-cumulantes_no_normalizados_q_nu}, it must be noticed that 
\begin{equation}\lim_{\nu\to\infty}C_{j,q_{(\nu)}}=\frac{\delta_{1j}}{2}\label{coeficiente}\end{equation}
so the dominant term in the scaling law \eqref{escalado_q-cumulantes_no_normalizados_q_nu} vanishes for $j>1$. Subdominant terms are also negligible, but in the limit $\nu\to\infty$ a dominant linear term arises so one recovers the extensive behaviour $\kappa_{j,1}^{(\infty)}\sim N$ advanced in \eqref{cumulantes_Boltzmann}.  

\begin{figure}
\centering\includegraphics[height=\linewidth,angle=-90,clip=]{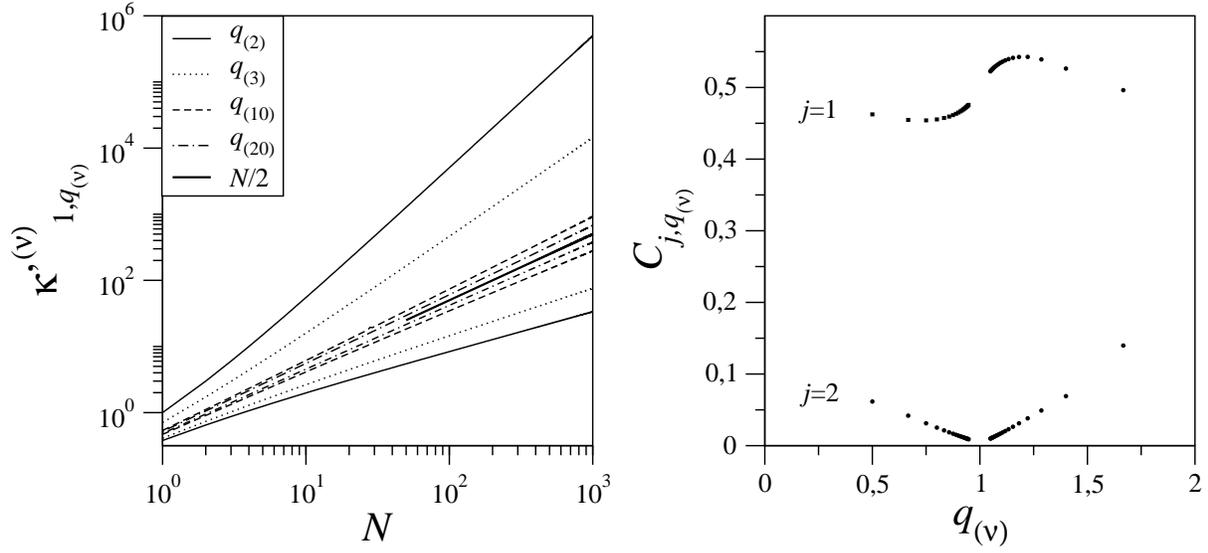}
\caption{(Left) Log-log plot of ${\kappa^\prime}^{(\nu)}_{1,q_{(\nu)}}={\mu^\prime}^{(\nu)}_{1,q_{(\nu)}}$ versus $N$ for conjugated pairs of triangles $\nu=2$, 3, 10 and 20. The limiting curve $\kappa^{(\infty)}_{1,1}=N/2$ is plotted as a guide to the eye. (Right) Proportionality coefficient $C_{j,q_{(\nu)}}$ in \eqref{escalado_q-cumulantes_no_normalizados_q_nu} for $j=1$ and 2 as a function of $q_{(\nu)}$. The expected limit \eqref{coeficiente} is obtained.
\label{fig_10}}
\end{figure}

Left panel of Fig.~\ref{fig_10} shows the order 1 $q_{(\nu)}-$cumulants in the Boltzmann-Gibbs limit $\nu\to\infty$. The predicted trend $\lim_{\nu\to\infty}\kappa^{\prime(\nu)}_{1,q_{(\nu)}}=\kappa^{\prime(\infty)}_{1,1}=\frac{N}{2}$ is clearly observed.  Right panel of Fig.~\ref{fig_10} shows the limit \eqref{coeficiente} for $j=1$ and 2. (Notice that Figs. \ref{fig_7} and \ref{fig_10} differ in the fact that for Fig. \ref{fig_7} we take the $\nu\to\infty$ limit for a $q-$cumulant with a fixed $q=3/2$ value whereas in Fig. \ref{fig_10} we take the same $\nu\to\infty$ limit but for the corresponding $q_{(\nu)}-$cumulant, so $q_{(\nu)}\to 1$). 

\section{Conclusion}
We have studied in detail the scaling with the system size of the $Q-$moments as well as the $q-$cumulants, ---normalized or not---, of the family of scale invariant triangles introduced in Section \ref{Scale_Invariant_Triangles}. We summarize the scaling laws that we have established in Table \ref{table_3}. The prefactors have been indicated whenever known. For the range of validity of expressions in the last column of the Table, see Section \ref{scaling_q-cumulants}. The scalings with $N$ of the $\nu<\infty$ column reduce to the $\nu \to\infty$ column for $q=1$ {\it except} for the unnormalized $q-$cumulants for $j> 1$. This unexpected feature (see Fig. 7), i.e., the fact that the $q \to 1$ and $N \to\infty$ limits do not commute, constitutes an interesting result of the present paper.

We believe that the present generalization of the cumulant expansion could be of interest for analytically discussing long-range-interacting systems (e.g., \cite{AntoniRuffo1995,AnteneodoTsallis1998}), for which moments above a given order typically diverge.    
\begin{table}
\begin{center}\begin{tabular}{cc|c|c|}\\\cline{3-4}
&&$\nu\to\infty\;(Q=q=1)$&$\nu<\infty\;(Q=j(q-1)+1)$\\\hline
\multicolumn{1}{|c|}{}&${\mu^\prime}^{(\nu)}_{1,Q}$&$N/2$&$\propto N^{(2-q)}$\\
\multicolumn{1}{|c|}{$j=1$}&$\gamma^{(\nu)}_{Q}$&1&$\propto N^{(1-q)}$\\
\multicolumn{1}{|c|}{}&$\mu^{(\nu)}_{1,Q}={\mu^\prime}^{(\nu)}_{1,Q}/\gamma^{(\nu)}_{Q}$&$N/2$&$N/2$\\
\multicolumn{1}{|c|}{}&${\kappa^\prime}^{(\nu)}_{1,q}$&$N/2$&$\propto N^{(2-q)}$\\\hline
\multicolumn{1}{|c|}{}&${\mu^\prime}^{(\nu)}_{2,Q}$&$\sim N^2/4$&$\propto N^{2(2-q)}$\\
\multicolumn{1}{|c|}{$j=2$}&$\gamma^{(\nu)}_{Q}$&1&$\propto N^{2(1-q)}$\\
\multicolumn{1}{|c|}{}&$\mu^{(\nu)}_{2,Q}={\mu^\prime}^{(\nu)}_{2,Q}/\gamma^{(\nu)}_{Q}$&$\sim N^2/4$&$\propto N^2$\\
\multicolumn{1}{|c|}{}&${\kappa^\prime}^{(\nu)}_{2,q}$&$N/4$&$\propto N^{2(2-q)}$\\\hline
\multicolumn{1}{|c|}{}&${\mu^\prime}^{(\nu)}_{j,Q}$&$\sim N^j/2^j$&$\propto N^{j(2-q)}$\\
\multicolumn{1}{|c|}{$j$}&$\gamma^{(\nu)}_{Q}$&1&$\propto N^{j(1-q)}$\\
\multicolumn{1}{|c|}{}&$\mu^{(\nu)}_{j,Q}={\mu^\prime}^{(\nu)}_{j,Q}/\gamma^{(\nu)}_{Q}$&$\sim N^j/2^j$&$\propto N^j$\\
\multicolumn{1}{|c|}{}&${\kappa^\prime}^{(\nu)}_{j,q}$&$\sim N$&$\propto N^{j(2-q)}$\\\hline
\end{tabular}\end{center}
\caption{Unnormalized $Q-$moments, normalization factors, ordinary $Q-$moments and unnormalized $q-$cumulants for de Boltzmann-Gibbs case (third column) and the family of triangles (fourth column) for $j=1$, 2 and the general case. In all cases, the symbols $\sim$ and $\propto$ are to be understood in the $N \to\infty$ limit. \label{table_3}}.
\end{table}

\appendix*
\section{}
For $q\in\mathbb R$, $q\neq 1$, the $q-$exponential function and its inverse, the $q-$logarithm function are defined as
\begin{align}
e_q^x&\equiv[1+(1-q)x]^\frac{1}{1-q}\\
\ln_qx&\equiv\dfrac{x^{1-q}-1}{1-q}
\end{align}
where the standard logarithm and exponential functions are recovered in the limit $q\to 1$. Their respective Taylor expansions are given by (see, for instance, the last of Refs.~\cite{libro})
\begin{equation}\begin{array}{rll}
e_q^x\!\!\!&=\displaystyle\sum_{n=0}^\infty\frac{e(q,n)}{n!}x^n;&\quad e(q,n)=\left\{\begin{array}{ll}1;&\quad n=0,1\\q(2q-1)\cdots((n-1)q-n+2);&\quad n>1\end{array}\right.\\\\
\ln_q(1+x)\!\!\!&=\displaystyle\sum_{n=1}^\infty\frac{(-1)^{n-1}}{n!}l(q,n)x^n;&\quad l(q,n)=\left\{\begin{array}{ll}1;&\quad n=1\\q(q+1)\cdots(q+n-2);&\quad n>1\end{array}\right.
\end{array}\end{equation}

\begin{acknowledgments}
The authors thank E.M.F. Curado for helpful comments, as well as partial financial support by CNPq and FAPERJ (Brazilian Agencies) and DGU-MEC (Spahish Ministry of Education) through Project PHB2007-0095-PC.  
\end{acknowledgments}

\bibliography{basename of .bib file}

\newpage


\end{document}